# Sequential Pitch Distributions for Raga Detection

**Vishwaas Narasinh Senthil Raja**

## ABSTRACT

Rāga is a fundamental melodic concept in Indian Art Music (IAM). It is characterized by complex patterns. All performances and compositions are based on the rāga framework. Rāga and tonic detection have been a long-standing research problem in the field of Music Information Retrieval. In this paper, we attempt to detect the rāga using a novel feature to extract sequential or temporal information from an audio sample. We call these Sequential Pitch Distributions (SPD), which are distributions taken over pitch values between two given pitch values over time. We also achieve state-of-the-art results on both Hindustani and Carnatic music rāga data sets with an accuracy of 99% and 88.13%, respectively. SPD gives a great boost in accuracy over a standard pitch distribution. The main goal of this paper, however, is to present an alternative approach to modeling the temporal aspects of the melody and thereby deducing the rāga.

## INTRODUCTION

Rāga is the central framework of Indian Art Music (IAM). The compositions and melodies are built on a given rāga. Each rāga is associated with a set of rules that define the flow, structure, and salience of notes (svaras). The melodies based on it are expected to adhere to those rules. A rāga will differ from another rāga either by having one or more differences in the set of notes, ascending and descending patterns, note salience, and melodic motifs.

Pitch estimation is the first step towards melody extraction and plays a very important role in musical signal processing. Rāga and tonic estimation typically require monophonic pitch tracking by the lead artist, although tonic estimation is usually enhanced by multi-track pitch estimation.
Another important step in rāga detection is estimating the tonic, as it sets the baseline for all the melody and notes [1]. Unlike several genres of music, typically Western classical music, that use absolute notes, IAM is based on relative notes. A rāga can be performed in any given tonic, and all the notes are defined with reference to this tonic [2]. A tonic is a perceptual quantity [3] which makes tonic detection subjective to different groups of people. The tonic corresponds to the note 'Sa' (the first note among the Indian solfege syllables [4][5] similar to the note 'Do' in Western classical music.
Rāga forms the basic framework for melodies and motifs in IAM [6]. A rāga can be thought of as a collection of melodic phrases and is typically performed in one tonic chosen by the lead artist throughout the performance. IAM typically involves improvisations of a rāga throughout a performance. A rāga can also involve continuous pitch variations in the form of Meend (gradual ascent or descent from one note to the other), Gamaks (embellishment that is placed on a note or in between two notes) [7], articulations, and different ascending and descending patterns [8], variations in the importance of each note. Unlike a pitch, a rāga is not a local feature, i.e., a piece of music has to be listened to for several seconds or minutes depending on the rāga progression and the listener's expertise in order to deduce the rāga. There have been various studies in the field of rāga detection, which is one of the most researched topics in IAM. Rāga modeling can be thought of as modeling a sequence of pitch values. Several rāga detection techniques are based on pitch histograms, and a handful of them model temporal structures. Temporal characteristics of a rāga are especially important in modeling the asymmetry between note transitions, which is important in distinguishing certain rāgas that share the same set of notes. And as there is no standard test data set available, all approaches to rāga recognition are evaluated on different datasets; a majority of them use a reduced number of classes, take a memory-based approach, or use a simple set of rāgas easy for classification.

## Literature Review

### Pitch Estimation

Several computational approaches towards pitch tracking have been studied, and some of the best results have been obtained by approaches that work on the frequency domain and utilize spectrograms and cepstrum [9]. PRAAT [10] is based on a normalized cross-correlation function and has been a popular choice. YIN [11] and an equivalent probabilistic version of it, pYIN [12] that use Hidden Markov Models have also been shown to give some of the best results. A more recent approach, CREPE [13] which we have used, is a highly efficient data-driven monophonic pitch estimation model based on Convolutional Neural Networks that works with raw data as opposed to candidate-generating functions like cepstrum or auto-correlation. CREPE [13] being one of the best models with an open-sourced Python implementation, is readily and easily available, motivating us to choose this over other pitch estimation approaches during real-time inference.

### Tonic Estimation

The tonic estimation is quite important, as tasks like rāga recognition [14][7][15][16], intonation analysis [17][18], and melodic motif analysis [19] are dependent on the right selection of the tonic. In the case of IAM, however [20], [21] uses melodic characteristics where the information regarding the amount of oscillation around each note called Gamak is considered for tonic estimation. This is especially useful in Carnatic, where the usage of Gamaks is quite strong. In a typical concert, a drone instrument, usually 'Tanpura', 'violin', or strings reinforces the tonic for the performer and the listener. [22] and [23] use this to enhance tonic estimation. [21] uses rāga information as a way of backtracking to find the tonic. Of all these multi-pitch approaches, [22] and [23] generally have the best accuracy.

### Rāga Detection

As one can imagine, a basic feature would be to extract the set of notes that describes the rāga, [20] does this by explicitly extracting the set of notes for rāgas. This is done for a handful of rāgas and this procedure does not extend to extracting melodic and temporal aspects of melody. As mentioned earlier, the salience of notes is also an important aspect of a rāga. A majority of studies have focused on extracting pitch distributions or similar variations of them [7][24][25]. This is usually done by taking the distribution of the 12 bin Pitch Class Distribution (PCD) or the fine-grained Pitch Distribution (PD) of each sample and comparing them with each other. This also captures the salience of pitches, which is important in rāga detection as one of the distinguishing features is Vadi and SamVadi (the first and second important note in a rāga). These approaches have shown to be fairly good in rāga classification tasks, a notable approach by [7] presents this approach on 23 rāgas with $91.5\%$ accuracy. The results are generally improved by taking a 120-bin PD, in contrast to a 12-bin PCD. [7], [26] Consider fine-grained pitch distribution to obtain better results. [7], [20] use probability density estimate using Gaussian-like kernels over pitch classes in a fine-grained pitch distribution, this is to capture the variation of each note around its mean as a way of capturing the Gamak information. This approach is shown to improve. In addition to taking pitch distributions [25] consider different statistics like mean, peaks, variance, kurtosis, and skewness, and report a better performance. All the above-mentioned approaches lack temporal or sequential information extraction of the melody, which is critical for many rāgas, An approach towards extracting sequential information is studied by [14] by forming dyads in the form of bi-grams along with PCD, the accuracy is said to improve with this, but they consider just 3 rāgas for classification. [27] extract ascending and descending features explicitly, which are used as templates, and claim to achieve 95% accuracy on $20$ rāgas. A few methods [28] use a pre-defined set of melodic patterns used for classification. The performance and generalisability of these methods are still questionable on a large set of rāgas. The current state-of-the-art results are obtained by Time Delayed Melodic Surface (TDMS) [16], where the authors create a matrix with two pitch values that are delayed by a time $\tau$ that is empirically set. The authors use K-Nearest Neighbours (KNN) to classify and Leave-One-Out-Cross-Validation (LOOCV) to evaluate. This makes (TDMS) [16] quite intuitive and simple to implement, and it captures temporal information. Our proposed method also tries to capture the temporal aspects using a completely different and novel approach. Our proposed method combines the Pitch Distribution aspect of earlier methods and the pitch transition aspect of TDMS [16]. This is done by the means of extracting the distribution of pitch values that lie between the transition of any two pitch values. This way, we also capture a sense of the direction of the transition. We call this feature Sequential Pitch Distributions (SPD), which are distributions taken between two pitch values. We also use an ensemble of K-Nearest-Neighbour models where each model captures different aspects of SPD. This method achieves better accuracy on both Hindustani and Carnatic datasets.

Evaluating our approach on all the above approaches is a rather difficult task due to different evaluation techniques, different data sets, and different environments. We nevertheless compare our accuracy with TDMS [16]. We use the same model and evaluation techniques as in TDMS for a fair comparison.
In this paper, we present a novel approach to extracting features that explain the sequential behavior of pitches and a K-Nearest Neighbours model that is trained on these features for rāga classification. The project, implemented in Python, along with the pre-trained models for Hindustani and Carnatic, is open source [1] and includes functions for easy evaluation and reproducibility.

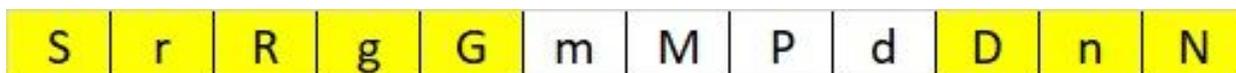

Fig 1. Highlighted pitches considered from Dha (D) to Ga (G) in the **positive** direction

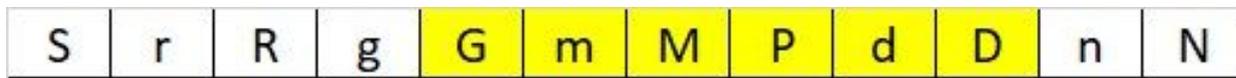

Fig 2. Highlighted pitches considered from Dha (D) to Ga (G) in the **negative** direction

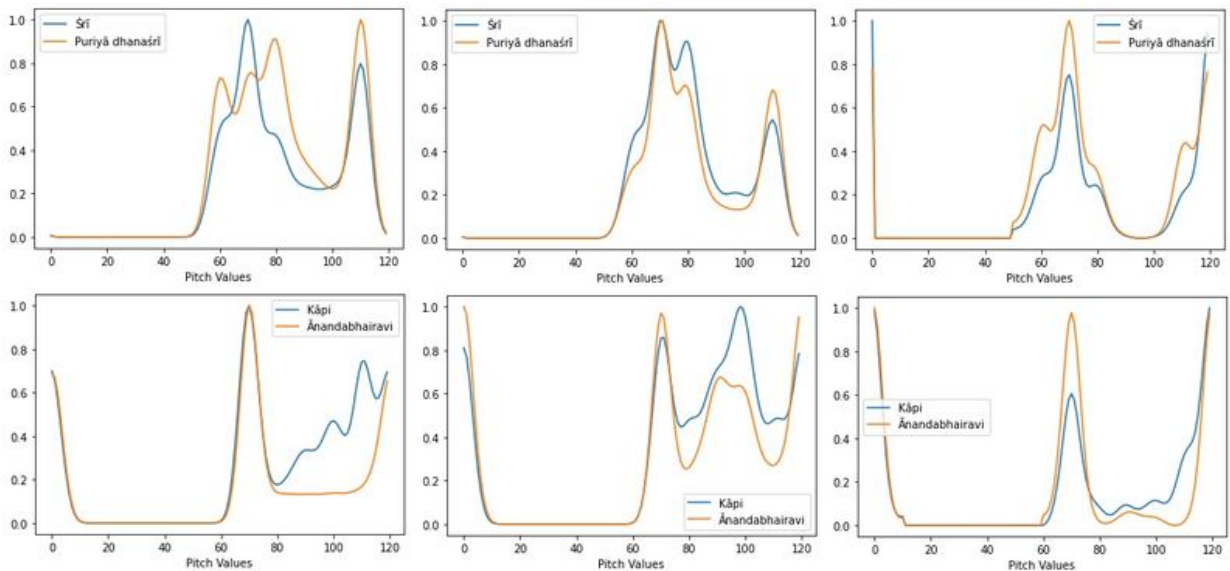

Fig 3. A subset of SPD of rāga Puriyā dhanaśrī and Śrī taken for a transition of notes from Ma' to Ni in the positive direction (top left), Ni to Ma' in the negative direction (top middle) and PD between notes Ma' and Ni (top right). A subset of SPD of rāga Kāpi and Ānandabhairavi taken for a transition of notes from Pa to Sa in positive direction order (Bottom left), Sa to Pa in negative direction order (Bottom middle) and PD between notes Pa and Sa (Bottom right)

## Model Overview

As rāga detection requires pitch and tonic estimation as prerequisites, we first obtain the pitch values from the annotated pitch files that are available in CompMusic [2] [29]. These files are used for training and testing. The pitch frequencies are converted to a 720-d vector spanning 6 octaves corresponding to pitches from C2 to B6, which amounts to 120 pitch values spanning an octave with 10 cents granularity. During run-time inference, however, we use CREPE [13] where the raw audio is sampled with a frame width of 1024 and a hop size of 30 ms. We compute the pitch values and apply Gaussian blur to each pitch value as described in CREPE [13]. Still, since rāga detection does not explicitly depend on octaves, we make it octave-independent by reducing the pitch vector by taking the sum over 6 octaves, resulting in a 120-d vector.

The tonic frequencies are also obtained from the annotations that are available in CompMusic. These frequencies are converted to a 120-d vector similar to pitch frequencies. The index corresponding to the max value in this vector is considered the tonic pitch value, to be used later to reorder the pitch values relative to this tonic.

Generally, for a new piece of audio containing a rāga (not already analyzed by CompMusic [29]), any pitch and tonic estimation system could be plugged into our rāga model to get rāga prediction.

## Rāga Detection with Sequential Pitch Distributions

Sequential Pitch distributions are obtained by taking the distributions of pitch values between any two given pitch values, namely $p_s$, and $p_e$. Given the pitch values of an audio sample, We first begin by searching the index of a start pitch value $p_s$ and then look for an end pitch value $p_e$. Both $p_s$ and $p_e$ correspond to the pitch values of the 12-chromatic scale. A transition from one pitch value to another can happen in two ways, namely in the positive and negative directions. An example of positive and negative features from Dha to Ga is shown in Fig. 1, 2. For simplicity, we have only shown pitches of the 12-chromatic scale; in practice, we use 120-bin fine-grained pitch values. Although it is true that when the pitches are folded into a single octave, the sense of direction of movement is lost, i.e., the transition from $p_s$ to $p_e$ may not have strictly ascending or descending pitches, it is nevertheless useful to consider it a feature. We also relax $p_s$ and $p_e$ by an amount $r = 4$, which ensures that pitch values that are within $\pm 40$ cents of $p_s$ and $p_e$ are also considered as start and end pitch values, respectively. The accuracies for different relaxations are given in Table. 2, this is quite intuitive, as in a practical scenario, a note may not be exactly on the standard chromatic scale but might be slightly off due to noise, pitch estimations, Gamak, or the musician itself. The case $r \geq 5$ is not possible in our setup as it results in an overlap between $p_s$ and $p_e$.

The below set of rules describes the positive feature of SPD.

A histogram of a sequence is taken when the sequence starts at index $i_s$ and ends at index $i_e$ such that

1. The pitch value at $i_s$ is $p_s \pm$ r cents.
2. The pitch value at $i_e$ is $p_e \pm$ r cents.
3. All pitch values between $i_s$ and $i_e$ belong to
$$\{p_s - r, p_s - r + 1, \ldots, p_e + r - 1, p_e + r\} \% 120$$

Here % indicates modulo operation.

Once valid $i_s$ and $i_e$ are calculated, the distributions of pitch values between them are extracted, this corresponds to the transition from $p_s$ to $p_e$. Any sequence that fails to obey the above rules for a given $p_s$ and $p_e$ is an invalid sequence and is discarded. In a given audio sample, there can be many $(i_s, i_e)$ pairs; distributions are extracted between each pair and then summed and normalized. In the case of the negative feature, the $r$ in the above 3 conditions is replaced by $-r$.

Since there can be cases where there are no valid sequences, a simple pitch distribution is extracted for the given audio sample instead. This proved to be extremely helpful in shorter audio samples, which improved the accuracy of short audio samples by $\sim 15\%$

With the relaxations in place, it turned out that it is quite sufficient if $p_s$ and $p_e$ take values corresponding to the standard chromatic scale, i.e., $p_s, p_e \in [0, 10, 20, 30, 40, 50, 60, 70, 80, 90, 100, 110]$.

Although $p_s$ and $p_e$ can take any pitch value in $[0, 120)$, taking all possible combinations of them will be computationally intensive. Taking all possible combinations of $p_s$ and $p_e$, and considering both positive and negative features, we get the SPD tensor $u$ of shape $(12, 12, 2, 120)$. For example, u[$p_s/10, p_e/10$, 0] is a 120-d feature that indicates a positive direction transition from $p_s$ to $p_e$; similarly, u[$p_s/10, p_e/10$, 1] is a 120-d feature that indicates a negative direction transition from $p_s$ to $p_e$. This is the main rāga feature. The accuracy of the models with this feature alone as input was 95% and 83%, for Hindustani and Carnatic traditions respectively. Furthermore, the feature dimension of the SPD tensor is quite high, which makes it challenging for a model to train. Hence, there is a need to split the SPD tensor into sub-features. This motivated a need for a musically informed feature set that would allow the models to focus on different sub-features and aspects of a rāga. Along with the main rāga feature, we derive two more sub-features, namely $v_1$ and $v_2$.

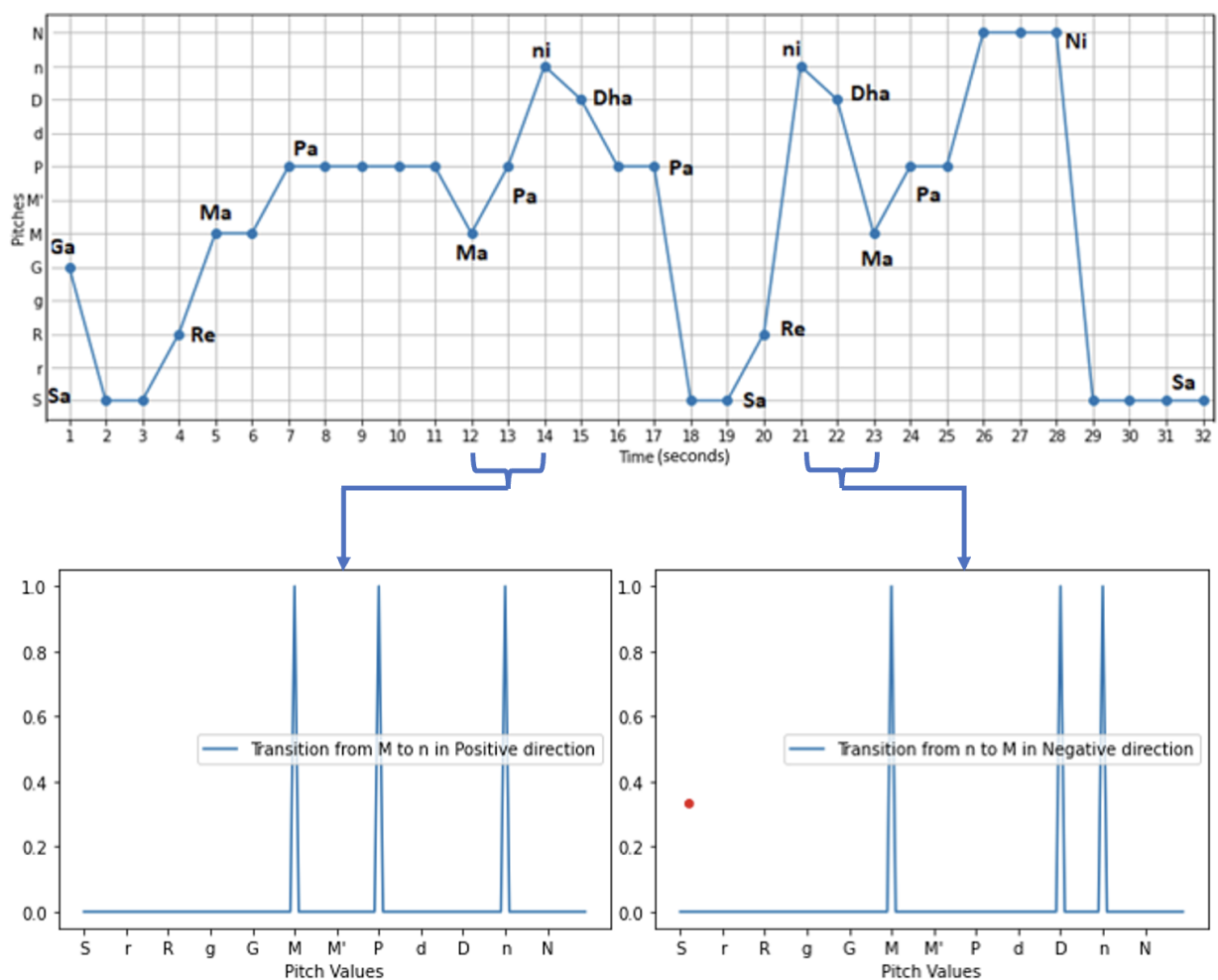

Fig 4. The pitch values for Baje Sargam song in rāga Des (Top) Example of a subset of SPD for a transition from M to n in the positive direction (left), the transition from n to M in the negative direction (middle), and pitch values for Baje Sargam song in

We define a modulo function

$$m(x) = x \% 12$$

and a function $v$

$$v(i,j,k) = u(i, m(j), k)$$

The split $v_1$ is obtained using

$$v_1(j,k) = [v(0,j,k), v(1, j+1, k), \ldots, v(11, j+11, k)]$$

The split $v_2$ is obtained using

$$v_2(i,k) = [v(i, i+1, k), v(i, i+2, k), \ldots, v(i, i+11, k)]$$

Where,

$$i, j \in [0, 11], j \neq 0$$

$$k \in [0, 1]$$

The tensors $v_1$ and $v_2$ have shapes $(11, 12, 2, 120)$ and $(12, 11, 2, 120)$ respectively. The sub-feature $v_1(j,k)$ extracts the subsets of SPD corresponding to the pitch values $(p_s, p_e)$ where $p_s$ and $p_e$ are separated by $10j$ cents. For example, $v_1(4, 0)$ extracts the subsets of SPD where $p_s$ and $p_e$ are separated by 40 cents. This allows a model trained on $v_1(4, 0)$ to learn the features that correspond to pitch values that differ by 40 cents. The sub-feature $v_2(i,k)$ extracts all subsets of SPD that have a start pitch value $p_s$ equal to $i$. This allows a model trained on $v_2(i,k)$ to learn the features that start with fixed pitch values.

| Tradition | TDMS (%) | SPD-KNN (%) | PD-KNN (%) |
|---|---|---|---|
| **Hindustani** | 97.67 | **99** | 90.66 |
| **Carnatic** | 86.7 | **88.13** | 72.33 |

Table 1. Accuracy comparison between TDMS, SPD, and Simple Pitch Distribution (PD) with the KNN model

An example of extracting a subset of SPD and the pitch values (obtained from CREPE) for a popular song, *Bhaje Sargam*, is shown in Figure [4](). Because of the restrictions imposed by the paper's dimensions in showing the actual pitches obtained, we have shown the pitch values of the main chorus of the song re-sampled at 1 sample per second and played at 60 beats per minute, and we have also rounded the pitch values to the nearest 12 standard chromatic scale pitch values for simplicity. The pitch values for this song have also been verified by experts in the field. The SPD subset of the transition taken from **M** to **n** in the positive direction is obtained by taking the distribution from time index 12 to 14 in Figure [4], similarly, the SPD subset of the transition taken from **n** to **M** in the negative direction is obtained by taking the distribution from time index 21 to 23 in Figure [4]. This is one of the SPD subsets that shows an important characteristic of the rāga showing the dissimilarities in the transitions. The distribution is not as smooth as the distributions shown in Figure [3] as the audio sample shown here has very few pitch values. Another example of the positive direction and negative direction features for the transitions between Ma' (60) and Ni (110) for the rāgas Puriyā dhanaśrī and Śrī is shown in Figure. [3] (top row). Although both rāgas share the same set of notes, the features clearly indicate dissimilar positive direction features and very similar negative direction features, as expected. Another example for rāgas Kāpi and Ānandabhairavi for the transitions between Pa (70) and Sa (0) is shown in Figure. [3] (bottom row). The positive and negative direction features show important differences between the rāgas even though both rāgas share the same notes. Also, note the positions of prominent peaks that correspond to the salient pitch values.

## K-Nearest-Neighbour Models

11 features of shape $(12, 2, 120)$ are derived from $v_1$ and 12 features of shape $(11, 2, 120)$ are derived from $v_2$, along with a simple pitch distribution and a complete SPD $u$ of shape $(12, 12, 2, 120)$, thus making 25 features overall. The accuracy of a single model trained on these 25 features as noted above, was 95% and 83%, for Hindustani and Carnatic traditions respectively. The reason for the relatively low accuracy is likely due to the high dimensionality of the features, which could make it difficult to model them with a single model. Hence, we construct 25 K-Nearest-Neighbour models, where each of the 25 models has been trained on a corresponding feature out of the total 25 features. The output of each of the models is a probability distribution over the target, and the output of the ensemble is of shape $(25, n_r)$, where $n_r$ is the number of ragas in a tradition. This is fed as input to a simple single-layer neural network whose output is a weighted average of these models' outputs. And inspired by TDMS, we use Bhattacharya distance $D_B$ as the distance metric for all KNN models with $k = 5$ for Hindustani and Carnatic.

$$D_B(m,n) = -\log\left(\sum \sqrt{f^m \cdot f^n}\right)$$

Where $f$ can be $u$, $v_1$, $v_2$, or pitch distribution

The accuracy for different values of $k$ is given in Table. [2] and we find that the accuracy remains more or less the same for different values of $k$. We also experimented with a few distance metrics, as shown in Table. [2] where $L_1$ is the Manhattan Distance. The KNN model enables us to compare our model in a LOOCV manner to TDMS.

# Dataset

The rāga datasets were requested from CompMusic. The dataset contains separate Hindustani and Carnatic datasets. The Hindustani datasets consist of 300 files for 30 rāgas, with 10 files for each rāga spanning 130 hours. The Carnatic datasets consist of 480 files for 40 rāgas, with 12 files for each rāga spanning 124 hours. The pitch files corresponding to them have a sampling rate of 4.44 ms, and the datasets also include the tonic frequencies for each file. The datasets are well balanced in terms of rāga, artists, and compositions and are completely vocal performances.

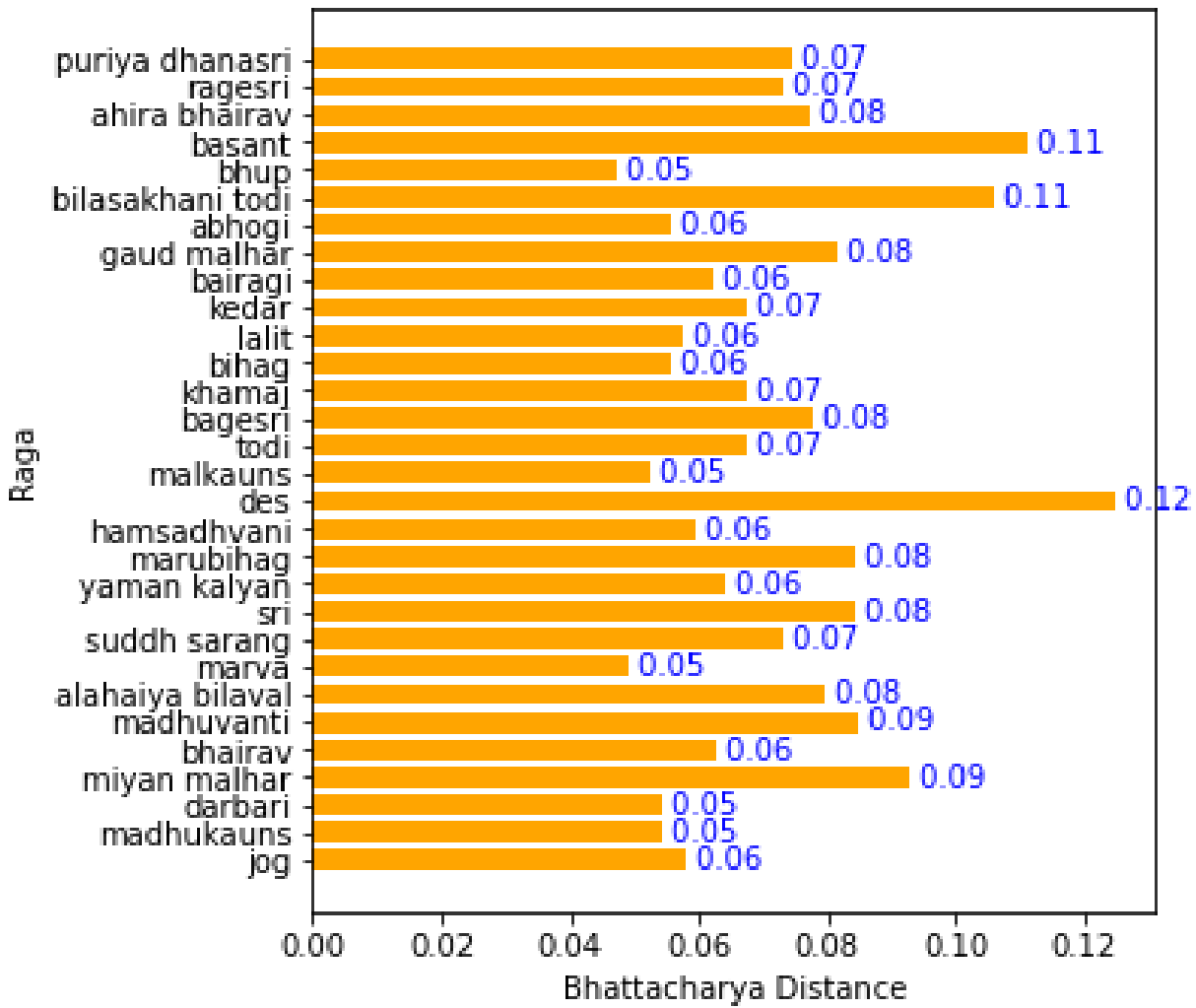

Fig 5. Bhattacharya Distance between Positive and Negative direction features of Hindustani rāgas

## Training and Evaluation Methodology

During the training process, SPD is calculated for each audio sample. Each of the 25 KNN models is trained in a LOOCV manner. The predictions by individual KNN models are combined in a weighted average, whose weights are calculated using a single-layer neural network. The SPD is cached for all the files, and we also cache the indexes $i_s$ and $i_e$ for faster training and evaluation. The evaluation metric is simple accuracy.

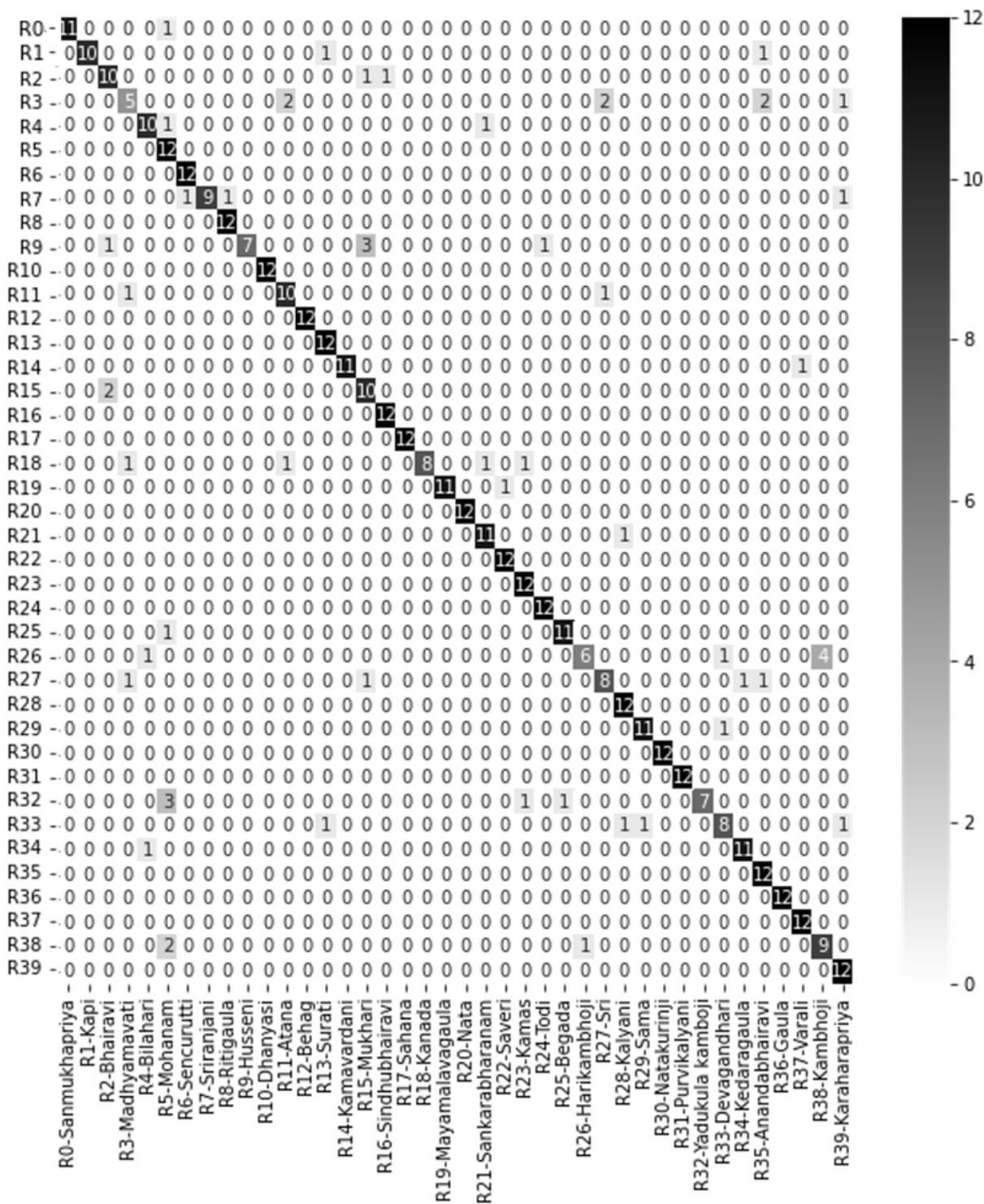

Fig 6. Confusion matrix of the predicted rāga labels obtained on the Carnatic rāga dataset. Shades of grey are mapped to the number of audio recordings.

| | **Nearest Neighbours (K)** | **Distance (D)** | **Relaxation (R)** |
| --- | --- | --- | --- |

| Tradition | 1 | 3 | 5 | 7 | $L_1$ | $D_B$ | 0 | 2 | 4 |
|---|---|---|---|---|---|---|---|---|---|
| **Hindustani** | 97.66 | 98.67 | **99** | 98.33 | 97.67 | **99** | 91.33 | 94.67 | **99** |
| **Carnatic** | 84.16 | 86.04 | **88.13** | 84.76 | 84.38 | **88.13** | 76.25 | 80.02 | **88.13** |

Table 2. Accuracy (%) for different nearest neighbors in the KNN model (with $D = L_1$ and R=4), distance metrics (With K=5 and R=4), and relaxation of start and end pitch values (with K=5 and $D = L_1$)

# Results

The accuracies for different approaches have been shown in Table. 1. The PD-KNN column contains the accuracy of Pitch Distribution with KNN (without SPD). The poorer performance on the Carnatic dataset is likely because of the relatively shorter audio files and higher number of classes compared to the Hindustani dataset.

We also calculate the Bhattacharya distance between positive and negative direction features of each rāga as shown in Figure. 5 for the Hindustani dataset. Some rāgas that have asymmetrical positive and negative directions are Des, Basant, and Miyan Malhar to name a few. We observe higher distances for rāgas that have asymmetrical positive and negative direction patterns, as expected.

We also analyze the predictions produced by our models. The confusion matrix of the predicted rāga labels for the Carnatic dataset is shown in Figure. 6, there are 57 incorrectly classified recordings. In general, we find that the confusions are in the rāgas in the sets {Kāmbhōji,Harikāmbhōji}, {Hussēnī,Mukhāri} and { Madhyamāvati, Śrī, Atāna}. These rāgas share a common set of notes and similar phrases [30]. For the Hindustani dataset, there are only 3 that are incorrectly classified. For 2 cases, the confusion is in the rāgas in the sets {Darbāri, Mārvā}, {Bāgēśrī, Khamāj} the error is in the estimation of the tonic.

# Conclusions and Future Scope

The method that we illustrated is not limited to Indian music but can be applied to any music genre that follows a tonic-rāga scheme. Also, SPD could be used for analyzing melodic phrases, motifs, and patterns to get a deeper understanding of a rāga.
One major improvement we see is the accuracy improvement for Carnatic-like traditions where the usage of Gamaks is quite strong, as noted above. We have considered $p_s$ and $p_e$ to be multiples of 10, which corresponds to the 12-pitch chromatic scale. Considering finer values for these could be another improvement, but this would require improving the computational complexity of SPD, which is currently quite intensive. Apart from accuracy improvements, optimizing the generation time for features during inference, and model improvements like extending to a deep learning model are some of the future scopes.

## Acknowledgments

I would like to pay our gratitude and respect to my father, Shri. Narasinh Hegde passed away due to COVID-19. He inspired me, taught me music, and helped me in several ways while working on this project. I also thank Sankalp Gulati for mentoring me and helping me improve this paper. Also, thanks to CompMusic for sharing the datasets.

## Footnotes

1. https://github.com/AnamikaConference/SPD_KNN ↩

2. https://compmusic.upf.edu/ ↩

## References

- Alekh, S. (2017). Automatic raga recognition in hindustani classical music. *arXiv Preprint arXiv:1708.02322*. ↩
- Bellur, A., Ishwar, V., Serra, X., & Murthy, H. A. (2012). A knowledge based signal processing approach to tonic identification in indian classical music. *Serra x, Rao p, Murthy h, Bozkurt b, Editors. Proceedings of the 2nd CompMusic Workshop; 2012 Jul 12-13; Istanbul, Turkey. Barcelona: Universitat Pompeu Fabra; 2012. P. 113-118*. ↩
- Boersma, P., & others. (1993). Accurate short-term analysis of the fundamental frequency and the harmonics-to-noise ratio of a sampled sound. *Proceedings of the Institute of Phonetic Sciences*, *17*, 97–110. ↩
- Chordia, P., & Rae, A. (2007). Raag recognition using pitch-class and pitch-class dyad distributions. *ISMIR*, 431–436. ↩
- Chordia, P., & Şentürk, S. (2013). Joint recognition of raag and tonic in north indian music. *Computer Music Journal*, *37*(3), 82–98. ↩
- De Cheveigné, A., & Kawahara, H. (2002). YIN, a fundamental frequency estimator for speech and music. *The Journal of the Acoustical Society of America*, *111*(4), 1917–1930. ↩
- Dighe, P., Karnick, H., & Raj, B. (2013). Swara histogram based structural analysis and identification of indian classical ragas. *ISMIR*, 35–40. ↩
- Gulati, S., Salamon, J., & Serra, X. (2012). A two-stage approach for tonic identification in indian art music. *Proceedings of the 2nd CompMusic Workshop; 2012 Jul 12-13; Istanbul, Turkey. Barcelona: Universitat Pompeu Fabra; 2012. P. 119-127*. ↩
- Gulati, S., Serrà Julià, J., Ganguli, K. K., Sentürk, S., & Serra, X. (2016). Time-delayed melody surfaces for rāga recognition. *Devaney j, Mandel MI, Turnbull d, Tzanetakis g, Editors. ISMIR 2016. Proceedings of the 17th International Society for Music Information Retrieval Conference; 2016 Aug 7-11; New York City (NY). [Canada]: ISMIR; 2016. P. 751-7*. ↩
- Gulati, S., Serra, J., Ishwar, V., Sentürk, S., & Serra, X. (2016). Phrase-based rāga recognition using vector space modeling. *2016 IEEE International Conference on Acoustics, Speech and Signal Processing*

*(ICASSP)*, 66–70.↩

- Henry, E. O. (1983). *The music of india: A scientific study*. ↩
- Kaup, S. J. (2011). *Cognitive processes for infering tonic*. ↩
- Kim, J. W., Salamon, J., Li, P., & Bello, J. P. (2018). Crepe: A convolutional representation for pitch estimation. *2018 IEEE International Conference on Acoustics, Speech and Signal Processing (ICASSP)*, 161–165. ↩
- Koduri, G. K., Gulati, S., Rao, P., & Serra, X. (2012). Rāga recognition based on pitch distribution methods. *Journal of New Music Research*, *41*(4), 337–350. ↩
- Koduri, G. K., Ishwar, V., Serrà, J., & Serra, X. (2014). Intonation analysis of rāgas in carnatic music. *Journal of New Music Research*, *43*(1), 72–93. ↩
- Koduri, G. K., Serrà Julià, J., & Serra, X. (2012). Characterization of intonation in carnatic music by parametrizing pitch histograms. *Gouyon f, Herrera p, Martins LG, Müller m. ISMIR 2012: Proceedings of the 13th International Society for Music Information Retrieval Conference; 2012 Oct 8-12; Porto, Portugal. Porto: FEUP Ediçoes, 2012.* ↩
- Krishna, T., & Ishwar, V. (2012). Carnatic music: Svara, gamaka, motif and raga identity. *Serra x, Rao p, Murthy h, Bozkurt b, Editors. Proceedings of the 2nd CompMusic Workshop; 2012 Jul 12-13; Istanbul, Turkey. Barcelona: Universitat Pompeu Fabra; 2012.* ↩
- Mauch, M., & Dixon, S. (2014). pYIN: A fundamental frequency estimator using probabilistic threshold distributions. *2014 Ieee International Conference on Acoustics, Speech and Signal Processing (Icassp)*, 659–663. ↩
- Noll, A. M. (1967). Cepstrum pitch determination. *The Journal of the Acoustical Society of America*, *41*(2), 293–309. ↩
- Ranjani, H., Arthi, S., & Sreenivas, T. (2011). Carnatic music analysis: Shadja, swara identification and raga verification in alapana using stochastic models. *2011 IEEE Workshop on Applications of Signal Processing to Audio and Acoustics (WASPAA)*, 29–32. ↩
- Rao, S., & Rao, P. (2014). An overview of hindustani music in the context of computational musicology. *Journal of New Music Research*, *43*(1), 24–33. ↩
- Ross, J. C., Vinutha, T., & Rao, P. (2012). Detecting melodic motifs from audio for hindustani classical music. *ISMIR*, 193–198. ↩
- Rowell, L. (2015). *Music and musical thought in early india*. University of Chicago Press. ↩
- Roychaudhuri, B. (2000). *The dictionary of hindustani classical music*. Motilal Banarsidass. ↩
- Salamon, J., Gulati, S., & Serra, X. (2012). A multipitch approach to tonic identification in indian classical music. *Gouyon f, Herrera p, Martins LG, Müller m. ISMIR 2012: Proceedings of the 13th International Society for Music Information Retrieval Conference; 2012 Oct 8-12; Porto, Portugal. Porto: FEUP Ediçoes; 2012.* ↩
- Serra, J., Koduri, G. K., Miron, M., & Serra, X. (2011). Assessing the tuning of sung indian classical music. *ISMIR*, 157–162. ↩

- Shetty, S., & Achary, K. (2009). Raga mining of indian music by extracting arohana-avarohana pattern. *International Journal of Recent Trends in Engineering*, *1*(1), 362. ↩
- Sridhar, R., & Geetha, T. (2009). Raga identification of carnatic music for music information retrieval. *International Journal of Recent Trends in Engineering*, *1*(1), 571. ↩
- Viswanathan, T., & Allen, M. H. (2004). *Music in south india : The karnatak concert tradition and beyond : Experiencing music expressing culture*. Oxford University Press. ↩
- Viswanathan, T., & Allen, M. H. (2004). *Music in south india: The karn† atak concert tradition and beyond: Experiencing music, expressing culture*. ↩